\documentclass
[twocolumn,english,aps,prl,superscriptaddress,showpacs,preprintnumbers]{revtex4}
\pdfoutput=1
\usepackage[latin9]{inputenc}
\usepackage{units}
\usepackage{amsmath}
\usepackage{graphicx}
\usepackage{amssymb}
\usepackage{nicefrac}
\usepackage{babel}
\usepackage{amsfonts}

\newcommand{\mvec}[1]{\vec #1}

\begin{document}
\title{Cavity optomechanics with ultra-high Q crystalline micro-resonators}

\author{J. Hofer}
\affiliation{Max Planck Institut f\"ur Quantenoptik, 85748 Garching, Germany}

\author{A. Schliesser}
\affiliation{Max Planck Institut f\"ur Quantenoptik, 85748 Garching, Germany}

\author{T.J. Kippenberg}
\email{tobias.kippenberg@epfl.ch}
\affiliation{Max Planck Institut f\"ur Quantenoptik, 85748 Garching, Germany}
\affiliation{Ecole Polytechnique F\'{e}d\'{e}rale de Lausanne, CH-1015, Lausanne, Switzerland}

\begin{abstract}
We present the first observation of optomechanical coupling in ultra-high $Q$ crystalline whispering-gallery-mode (WGM) resonators. The high purity of the crystalline material enables optical quality factors in excess of $10^{10}$ and finesse exceeding $10^{6}$. Simultaneously, mechanical quality factors greater than $10^5$ are obtained, still limited by clamping losses. Compared to previously demonstrated cylindrical resonators, the effective mass of the mechanical modes can be dramatically reduced by the fabrication of CaF$_2$ microdisc resonators. Optical displacement monitoring at the $10^{-18}$ m/$\sqrt{\mathrm{Hz}}$-level reveals mechanical radial modes at frequencies up to 20 $\mathrm{MHz}$, corresponding to unprecedented sideband factors ($>100$). Together with the weak intrinsic mechanical damping in crystalline materials, such high sindeband factors render crystalline WGM micro-resonators promising for backaction evading measurements, resolved sideband cooling or optomechanical normal mode splitting. Moreover, these resonators can operate in a regime where optomechanical Brillouin lasing can become accessible. 

\end{abstract}
\maketitle

Optical interferometers with suspended mirrors have traditionally been key elements in gravitational wave detectors. Implemented at a mesoscopic scale ($\lesssim 1$ cm), micro- and nano-scale physical systems coupling optical and mechanical degrees of freedom may allow studying optomechanical coupling at the quantum level \cite{Kippenberg2008,Schwab2005}. In particular, recent experiments have aimed towards the observation of measurement quantum backaction \cite{Verlot2009,Anetsberger2009} and radiation-pressure cooling of a mesoscopic mechanical oscillator to its quantum ground state \cite{Schliesser2009,Park2009,Groeblacher2009b}. However, inevitable coupling of the system to its environment severely impedes such studies. For example, thermal fluctuations associated with mechanical dissipation mask the signatures of quantum backaction \cite{Tittonen1999,Verlot2009,Anetsberger2009}, but also constitute a mechanism competing with radiation-pressure cooling \cite{Schliesser2009,Park2009,Groeblacher2009b}. Optical losses, on the other hand, destroy potential quantum correlations \cite{Verlot2009}, give rise to heating due to the absorbed photons \cite{Schliesser2009}, and preclude reaching the important resolved-sideband regime \cite{Schliesser2008}. Most of the various optomechanical systems investigated recently are based on amorphous material such as $\mathrm{SiO}_{2}$ \cite{Schliesser2008,Anetsberger2008} or $\mathrm{Si}_3\mathrm{N}_4$ \cite{Thompson2008,Wilson2009,Groeblacher2009b}. The absence of microscopic order makes these materials prone to mechanical losses due to coupling of strain fields to two-level systems (TLS) \cite{Pohl1970,Pohl2002}, particularly severe at cryogenic temperatures \cite{Arcizet2009}. Strained $\mathrm{Si}_3\mathrm{N}_4$ oscillators have achieved higher mechanical quality factors $Q_{\mathrm{m}}\approx 1\times10^{7}$, however, optical finesse ($\mathcal{F}$) has been limited to values below $10^4$ \cite{Thompson2008,Wilson2009}, although the imaginary part of the refractive index can be lower than $10^{-5}$ \cite{Wilson2009}. Crystalline materials, in contrast, do not suffer from such restrictions at all. In fact, the best optical resonators available today are whispering-gallery mode (WGM) resonators made from CaF$_{2}$, featuring $\mathcal{F}$ up to $10^{7}$ \cite{Savchenkov2007,Ilchenko2004,Grudinin2006a,Grudinin2006b}. At the same time, due to their long range order, crystalline materials are ideally free from TLS leading to ultra-high mechanical quality factors. Indeed, remarkably high values up to $4\times 10^{9}$ have been reported for a 1-MHz bulk crystalline quartz oscillator at a temperature of 2 K \cite{Smagin1974}. \newline
Here, we combine the generally favorable properties of monolithic WGM optomechanical systems in terms of optomechanical coupling strength \cite{Kippenberg2005b,Schliesser2008b}, stability and 
\begin{figure}[h]
\centering  
\includegraphics[width=.44\textwidth]{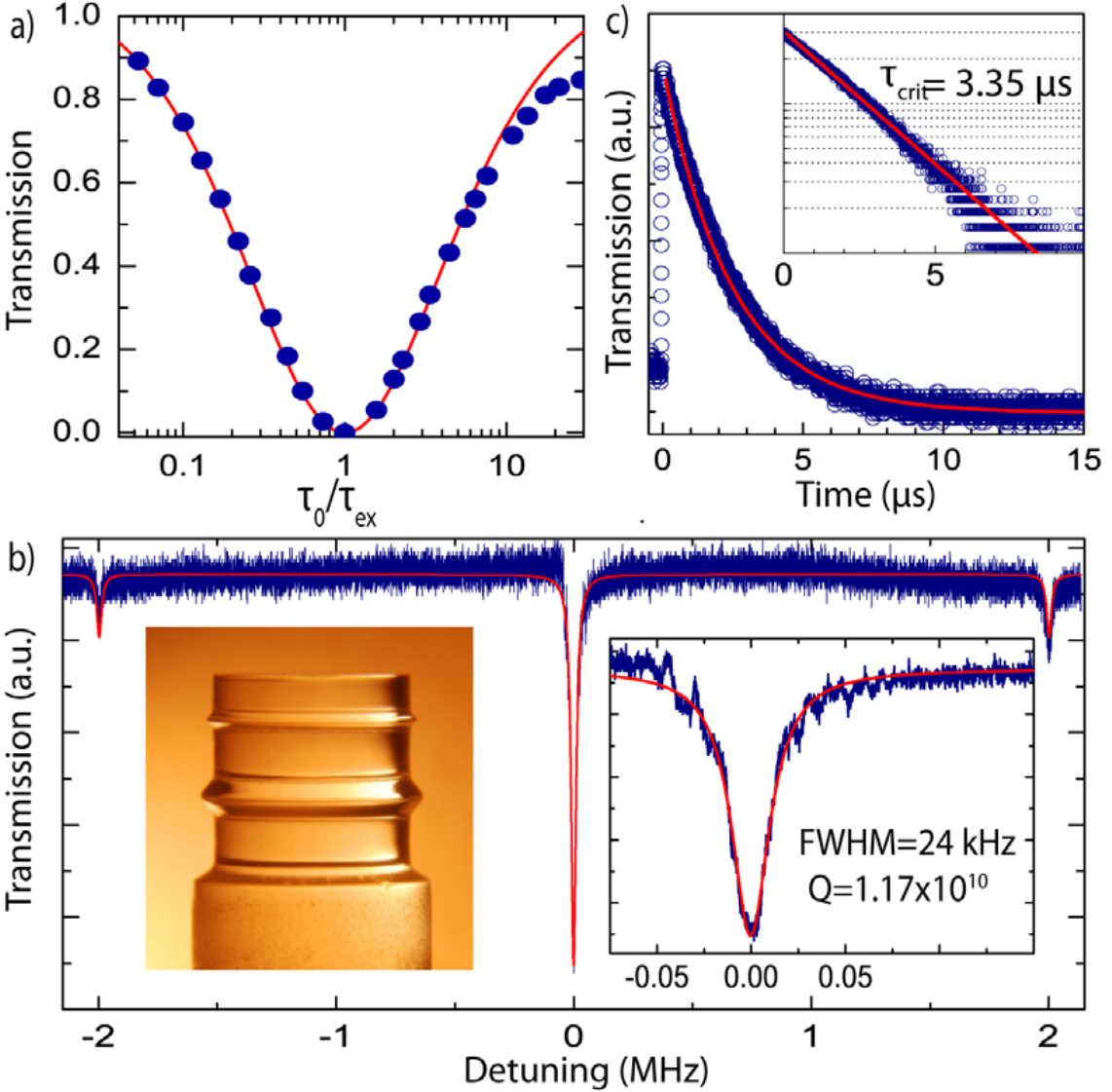} 
\caption{a) Transmission on resonance vs. coupling parameter $\tau_{0}/\tau_{\mathrm{ex}}$ while approaching a tapered fiber to a CaF$_{2}$ cavity. The data (blue points) show that ideal coupling behavior (red line) is possible up to a coupling factor $\tau_{0}/\tau_{\mathrm{ex}} \approx 8$. b) Measurement of the cavity linewidth with calibration peaks at 2 $\mathrm{MHz}$ detuning and Lorentzian fit (red line). Inset shows a diamond-turned resonator made of CaF$_{2}$. c) Ringdown measurement (blue points) of the same resonator and exponential fit (red line). The measured lifetime of $\tau_{\mathrm{crit}}=3.35$ $\mu$s corresponds to an intrinsic quality factor of $Q_{0}=1.18\times10^{10}$.}
\label{fig1}
\end{figure}
cryogenic compatibility \cite{Schliesser2009} with a crystalline material. 
High-Q mechanical modes are observed in cylinder microresonators and their motional mass determined. By shaping thin disk resonators from the same material, radial breathing modes are observed with sub milli-gram motional mass. Ultra-long photon storage times $\kappa^{-1}$ and high mechanical frequencies $\Omega_{\mathrm{m}}$ enable remarkably large sideband factors $\Omega_{\mathrm{m}}/\kappa$ in these systems, up to unprecedented values exceeding 100. In Fabry-Perot type cavities, such sideband factors could only be achieved with long cavities, resulting in significantly lower optomechanical coupling. The large sideband factors achieved here are of interest for a variety of experimental schemes, such as sideband cooling \cite{Schliesser2008}, back action evading measurements \cite{Braginsky1992} as well as strong optomechanical coupling \cite{Groeblacher2009}. Moreover the demonstrated systems are the first to operate in a novel regime in which mechanical dissipation can exceed optical dissipation, enabling observation of optomechanical Brillouin lasing \cite{Grudinin2010}.

Following pioneering work \cite{Grudinin2006a,Grudinin2006b} we fabricate crystalline WGM resonators from purest available CaF$_{2}$ or MgF$_{2}$ raw material. In brief, small cylindrical and disc preforms were drilled out of crystalline blanks and mounted onto an air-bearing spindle, which provides extremely low rotational imprecision ($<100$ nm). The preforms were shaped using diamond turning and, to minimize surface scattering, a polishing procedure with diamond abrasive of successively decreasing grit size (down to an average size of 25 nm) was applied.

As demonstrated below for the first time, highly efficient evanescent coupling to CaF$_{2}$ and MgF$_{2}$ resonators can be achieved by the use of tapered optical fibers. Tapered fibers conveniently provide single-mode input and output as well as tunable coupling \cite{Spillane2003}. Depending on the size of the resonator and the used wavelength, phase matching is achieved by adjusting the taper waist radius. For the used CaF$_{2}$ resonators of radius $R=800$ $\mu$m and 2 mm, the optimum taper waist corresponds to 1.15 and 1.25 $\mu$m radius, respectively. As an example, Fig. 1a shows coupling to a CaF$_{2}$ resonator of 800 $\mu$m radius possessing an intrinsic optical quality factor $Q_{\mathrm{0}}=\tau_{\mathrm{0}}\omega=1.4\times10^{9}$, where $\tau_{\mathrm{0}}$ is the intrinsic photon lifetime and $\omega$ the optical carrier frequency. The coupling to the waveguide can be described by the dimensionless parameter $\tau_{\mathrm{0}}/\tau_{\mathrm{ex}}$, where $\tau_{\mathrm{ex}}^{-1}$ reflects the photon loss rate due to coupling to the waveguide and the resulting linewidth $\kappa$ is given by $\kappa = \tau_{\mathrm{0}}^{-1} + \tau_{\mathrm{ex}}^{-1}$. The data clearly show that critical ($\tau_{\mathrm{0}} = \tau_{\mathrm{ex}}$) and strong overcoupling up to $\tau_{\mathrm{0}}/\tau_{\mathrm{ex}} = 30$ are possible, however, coupling deviates from ideality \cite{Spillane2003} for $\tau_{\mathrm{0}}/\tau_{\mathrm{ex}}>8$. Operation in the highly overcoupled regime---demonstrated here for the first time for crystalline resonators---is a key prerequisite for e.g sensitive detection of mechanical motion \cite{Schliesser2008b}, efficient optical comb generation \cite{DelHaye2007,Savchenkov2008}, and is critical in experiments involving quantum correlations \cite{Verlot2009}.
\begin{figure}[h]
\centering  \includegraphics[width=.4\textwidth]{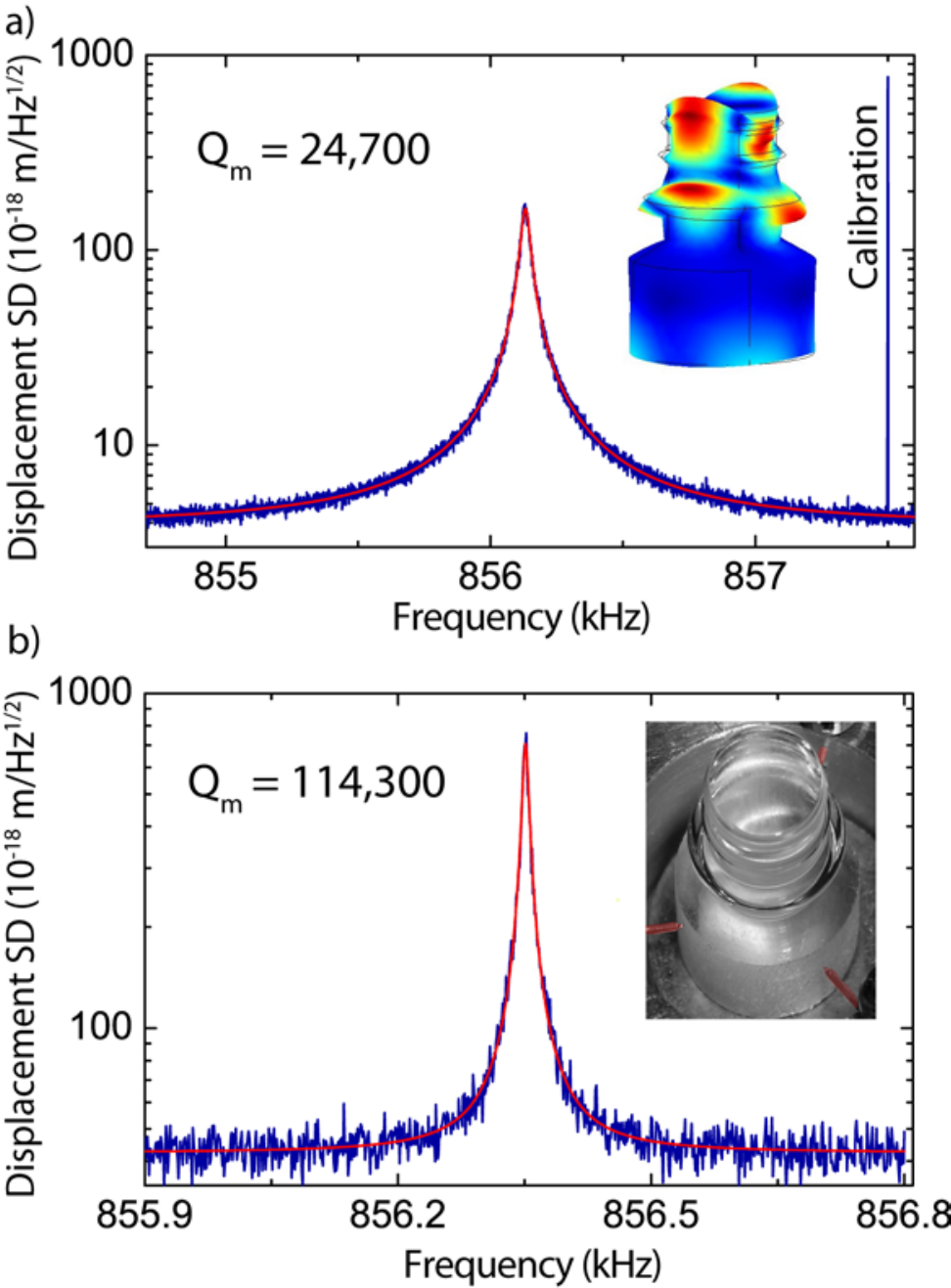}
\caption{a) Mechanical mode of a cylindrical cavity measured free-standing at atmospheric pressure. Blue line are data, red line a Lorentzian fit. For absolute displacement calibration, the laser was phase modulated at 857.5 $\mathrm{kHz}$. Inset shows the typical displacement pattern of a mode in this frequency range. b) The quality factor of the same mechanical mode measured at low pressure and with optimized clamping increased to 114,300. Inset illustrates the mounting of the resonator. It is clamped by four tungsten tips, three sidewise (red shaded) and one from below.}
\label{fig2}
\end{figure}

The resonator quality factor was inferred from linewidth measurements and cavity ringdown experiments. 
The linewidth was measured with a scanning Nd:YAG laser at 1064 nm. In the undercoupled regime ($\tau_{\mathrm{ex}}\gg\tau_{\mathrm{0}}$) the highest measured Q-factor was $1.17\times10^{10}$. In addition, the Q-factor was independently measured using cavity ringdown which is principally insensitive to thermal effects and not limited by the laser linewidth. Data from a typical ringdown measurement are shown in Fig. 1c. The total lifetime $\tau_{\mathrm{crit}}$ at critical coupling was in this case 3.35 $\mu$s. The corresponding intrinsic cavity $Q_{\mathrm{0}} = 1.18\times10^{10}$ is in excellent agreement with the linewidth measurements and corresponds to an intrinsic finesse $\mathcal{F} = 770,000$. The highest value attained in such a measurement was $\mathcal{F} =$ 1.1 million for a CaF$_{2}$ resonator of 550 $\mu$m in radius, constituting the highest reported finesse in an optomechanical system to date. Similar values were measured with MgF$_{2}$ resonators. 

WGM resonators possess inherent optomechanical coupling of the optical mode with structural mechanical resonances which displace the resonator's boundaries \cite{Kippenberg2005b,Schliesser2008b}. We studied this ponderomotive interaction in several different resonator geometries made of CaF$_{2}$ and MgF$_{2}$. 
We first discuss the results of a typical mm-scale CaF$_{2}$ cylindrical resonator, 8 mm in height and 2 mm radius, similar to the geometry of the resonator shown in Fig. 1b. To detect the mechanical modes' Brownian motion, a weak beam from a cw Ti:Sapphire laser was coupled to a WGM using a tapered optical fiber, and locked to the side of an optical resonance's fringe. Thermal fluctuations of the cavity radius cause changes in the optical path-length, which imprint themselves into a modulation of the power transmitted past the cavity. Spectral analysis of the transmitted power thus allows extracting the mechanical frequency and dissipation rate $\Gamma_{\mathrm{m}} = \Omega_{\mathrm{m}}/Q_{\mathrm{m}}$. Alternatively, to achieve a quantum-limited displacement sensitivity \cite{Schliesser2008b,Arcizet2006} a phase sensitive polarization spectroscopy scheme was used. Using these methods, more than 20 mechanical modes with frequencies ranging from ca.\ 500 $\mathrm{kHz}$ to 2 $\mathrm{MHz}$ were observed. For a freely standing cylinder we obtain $Q_{\mathrm{m}} \sim 20,000$. In order to reduce the influence of clamping losses and gas friction, the resonator was mounted with four sharp tungsten tips (apex $\approx$ 1 $\mu$m) and measured in a low pressure environment ($<10$ mbar). As an example a mechanical mode with an eigenfrequency of $856$ $\mathrm{kHz}$ is shown in Fig. 2, at atmospheric and at low pressure (with $Q_{\mathrm{m}}$ of $24,700$ and $114,300$, respectively). In this configuration several modes had Q-factors exceeding $10^{5}$ with a maximum $Q_{\mathrm{m}}$ of 136,000 at a frequency of 1.1 $\mathrm{MHz}$. Note that this corresponds to a Q-frequency product greater than $10^{11}$ at room temperature. Due to the strong dependence of the measured $Q_{\mathrm{m}}$ on the device's clamping, we expect that significantly higher $Q_{\mathrm{m}}$ can be achieved by optimizing the clamping. 

To fully assess the performance of an optomechanical system, it is also necessary to quantify the optomechanical coupling strength, which is usually expressed in terms of the (mutually dependent) coupling parameter $g_0$ and effective mass $m_{\mathrm{eff}}$. The former indicates the differential frequency shift for a given displacement $x$ via $g_0 =\partial \omega/\partial x$. In complex three-dimensional structures such as those employed here, however, the definition of $x$ (and therefore $g_0$) is indeed arbitrary, a fact that has received little attention so far. Physically, the mechanical modes (enumerated in the following with an index n) are characterized by a three-dimensional displacement field $\vec u_{\mathrm{n}}(\mvec r)$, which induces a relative optical frequency shift of 
\begin{equation}
  \frac{\delta \omega}{\omega}\approx\frac{\int|\mvec E(\mvec r)|^2\,\mvec \nabla \varepsilon(\mvec r)\cdot \mvec u(\mvec r) d^3 r}{2\int|\mvec E(\mvec r)|^2\, \varepsilon(\mvec r) d^3 r}.
\end{equation}
in the case of weak index contrast (where field discontinuities are small) \cite{Johnson2002}, where $\vec u(\mvec r) = \sum_{\mathrm{n}} \vec u_{\mathrm{n}}(\mvec r)$, and $\vec E(\mvec r)$ and $\varepsilon(\mvec r)$ are the electric field and dielectric constant, respectively. For a given choice of $g_0$, the equivalent scalar displacement can be written as $x\equiv \delta \omega/g_0=  \langle \mvec w, \mvec u\rangle \equiv \int \mvec w(\mvec r)\cdot \mvec u(\mvec r)d^3r$, where the weighting function $\mvec w(\mvec r)$ is introduced such that eq.\ (1) is satisfied. Note that, as a consequence, the magnitude of $\mvec w(\mvec r)$ scales inversely with $g_0$. The mapping of the displacement vector $\mvec u_{\mathrm{n}}(\mvec r)$ to the scalar $x$ necessitates  the introduction of an effective mass $m_{\mathrm{n}}$ for each mode \cite{Gillespie1995, Pinard1999}. From the requirement that the potential energy upon a displacement $x_{\mathrm{n}}$ of the n-th mode is given by $U_{\mathrm{n}}=\frac{1}{2}m_{\mathrm{n}} \Omega_{\mathrm{n}}^2 x_{\mathrm{n}}^2$ it can be shown \cite{Pinard1999} that the effective mass is given by
\begin{equation}
  m_{\mathrm{n}}=\frac{\int \rho(\mvec r)  |\mvec u_{\mathrm{n}}(\mvec r)|^2 d^3r}{ \langle \mvec w, \mvec u_{\mathrm{n}}\rangle^2}
\end{equation}
evidently scaling quadratic in the choice of coupling constant $g_0$ ($\rho(\mvec r)$ is the density). Note that the optomechanical coupling Hamiltonian $\hat H_{\mathrm{int}}=\hbar \, g_0 \, \hat a_{\mathrm{o}}^{\dagger} \hat a_{\mathrm{o}} \, x_{\mathrm{zpf}}\, (\hat a_{\mathrm{m}}^{\dagger} + \hat a_{\mathrm{m}})$, where $\hat a_{\mathrm{o}}$ ($\hat a_{\mathrm{o}}^{\dagger}$) and $\hat a_{\mathrm{m}}$ ($\hat a_{\mathrm{m}}^{\dagger}$) are the optical and mechanical annihilation (creation) operators, respectively, is independent from the choice of $g_0$, as $x_{\mathrm{zpf}}=\sqrt{\hbar/(2 m_{\mathrm{n}} \Omega_{\mathrm{n}})}$. In the case of WGM optomechanical systems, a natural choice for the optomechanical coupling is given by $g_0=-\omega/R$, mapping the displacement fields to an effective change in the radius of the entire structure. From room temperature measurements, we can then experimentally derive the effective mass from the calibrated \cite{Schliesser2009} displacement spectra $\bar S_{\mathrm{xx}}(\Omega)$ using the relation $\bar S_{\mathrm{xx}}(\Omega_{\mathrm{n}})=2 k_\mathrm{B} T /(m_{\mathrm{n}} \Omega_{\mathrm{n}}^2\Gamma_{\mathrm{n}}$). 

The effective masses measured in this manner on the cylindrical samples are still comparably high ($m_{\mathrm{eff}}=50$ mg). For optomechanical experiments this value is required to be substantially reduced. We demonstrate that the effective mass can be dramatically reduced by fabricating disc shaped crystalline resonators. Indeed, for a CaF$_{2}$ resonator of 100 $\mu$m thickness and 800 $\mu$m radius, a reduction of over two orders of magnitude in effective mass is attained and higher optomechanical coupling $\left|g_{0}\right|=350$ MHz/nm achieved. The magnitude of the intrinsic optical Q-factor of the disc was $1.4\times10^{9}$ and mechanical modes were measured with a Nd:YAG laser locked to the side of a cavity resonance, which allowed a shot-noise limited displacement sensitivity at the level of $1\times10^{-18}\mathrm{m}/\sqrt{\mathrm{Hz}}$ above 6 $\mathrm{MHz}$. Below 6 $\mathrm{MHz}$ the sensitivity was limited by classical laser noise, which was characterized using an independent fiber loop cavity. In a first measurement the CaF$_{2}$ disc was waxed on a metal holder that was used in the fabrication process. Besides the high-$Q$ modes ($Q_{\mathrm{m}}$ up to 15,000), the most prominent feature in the noise spectrum as shown in Fig. 3a are three broad Lorentzian peaks which can be attributed to different orders of radial breathing modes (RBMs, $Q_{\mathrm{m}} < 100$) with effective masses from 400 $\mu$g to 700 $\mu$g. 
\begin{figure}[h]
\centering  \includegraphics[width=.43\textwidth]{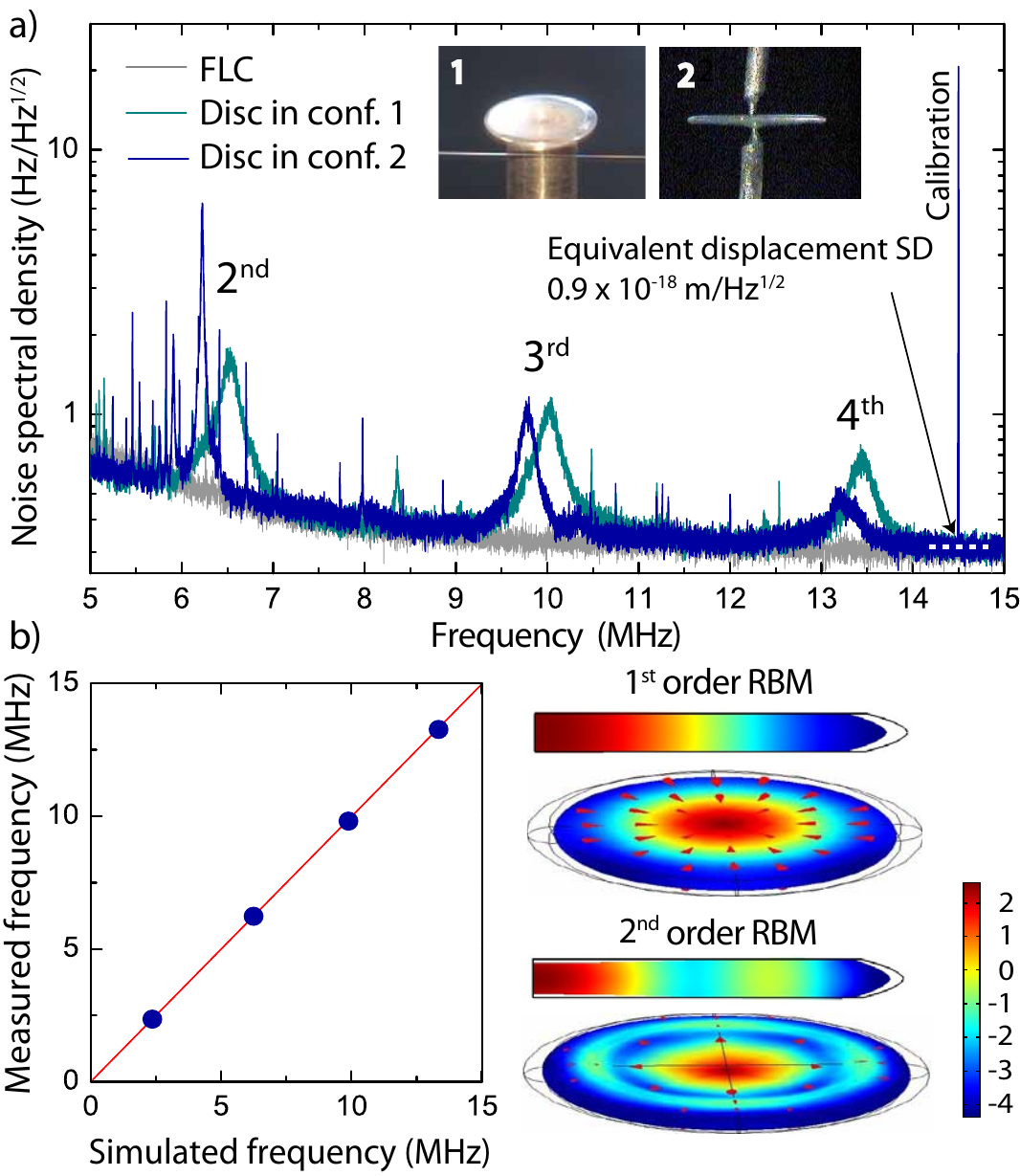} \caption{a) Noise spectral density of a CaF$_{2}$ microdisc waxed on a metal holder (configuration 1), clamped between two tungsten tips (configuration 2), and a fiber loop cavity (FLC) measured with a Nd:YAG laser. The disc exhibits several orders of mechanical RBMs up to 20 MHz (second to fourth order are shown). b) For the case of central clamping, measurement and simulation of first to fourth order RBM frequencies show excellent agreement. The increase in modal displacement at the support tip for higher order RBMs explains the measured dependence of $Q_{\mathrm{m}}$ on the RBM order (from $Q_{\mathrm{m}}$ = 2500 to $Q_{\mathrm{m}}$ = 40). Color code represents radial displacement (arbitrary units).}
\label{fig3}
\end{figure}
Note that in addition to the low mass the frequency of these modes falls in the range $>10$ MHz and therefore provides hereto unprecedented sideband factors ($>100$). To reduce coupling of the RBMs to the environment, the disc was clamped centrally between two sharp tungsten tips, which led to a substantial increase of the Q-factor, most distinct for lower-order RBMs. 
While for the 1$^{\mathrm{st}}$ order RBM a quality factor of $2,500$ was measured in this configuration, it was $300, 70$ and $40$ for 2$^{\mathrm{nd}}$, 3$^{\mathrm{rd}}$ and 4$^{\mathrm{th}}$ order, respectively, limited by clamping losses as explained below.

To verify the nature of the observed modes we used finite element modeling (COMSOL Multiphysics). CaF$_{2}$ is a cubic crystal and described by three independent elastic constants. Taking into account the crystalline orientation (symmetry axis of the resonator parallel to $\mathrm{[111]}$), and appropriate boundary conditions, the mechanical eigenfrequencies, mode energies, and stress and strain fields were inferred from the simulation. Three-dimensional simulations yield very accurate values for the mode frequencies but also a dense mode spectrum ($\sim$ 20 modes per $\mathrm{MHz}$). Most of the modes with small masses can already be identified in a 2-dimensional model exploiting the cylindrical symmetry of the boundary conditions. For the case of central clamping between two sharp tips, a stress free boundary condition was applied and matching of measured and simulated frequencies of first to fourth order RBM with an accuracy of better than $\pm 1 \%$ was achieved. The simulated mechanical displacement fields $\mvec u_{\mathrm{n}}(\mvec r)$ also provide a way to numerically estimate the effective mass of radially symmetric modes: Approximating the weighting function by a circle of radius $R$, we extract the radial displacement $x_{\mathrm{n}}$ of a given radially symmetric mode at the rim of the resonator and directly solve for the effective mass using the mode's known resonance frequency and energy $U_{\mathrm{n}}$. Indeed, measured and simulated RBM effective masses show good agreement ($\pm 10 \%$). FEM was also used to identify  modal patterns and the origin of clamping losses. As stated above, the Q-factor in the case of central clamping strongly decreases with increasing order of the RBM. With the FEM simulation this effect can be clearly understood by the rising axial displacement at the central point for higher order RBMs, which leads to stronger coupling and dissipation to the environment via the tungsten support tip. FEM simulations can be employed to engineer higher-$Q$ modes and clamping geometries \cite{Anetsberger2008}. 

In summary, we have for the first time observed and characterized optomechanical coupling in crystalline WGM resonators, possessing both high optical and mechanical Q-factors. Moreover unprecedented sideband factors ($>100$) are attained, and means to reduce effective mass and clamping losses demonstrated using a novel disc shaped resonator. Employing highly efficient tapered fiber coupling and a shot-noise limited detection scheme, radial breathing modes with high frequency ($>10$ $\mathrm{MHz}$) are observed. The reported mechanical quality factors are far from material-limited; as a reference value $Q_{\mathrm{m}} = 3\times10^{8}$ was measured in CaF$_{2}$ at a frequency of 40 $\mathrm{kHz}$ and 60 K \cite{Nawrodt2007}. Nonetheless, the systems presented here allow the observation of radiation-pressure induced parametric instability \cite{Kippenberg2005b} and compare very favorably to systems suggested to observe optomechanically induced quantum correlations \cite{Verlot2009} exhibiting a 1000-times higher mass.
The strong increase of the quality factor at low temperatures is in stark contrast to the universal degradation of $Q_{\mathrm{m}}$ in amorphous glass \cite{Pohl2002}, limiting optomechanical cooling experiments \cite{Schliesser2009, Park2009}. The resonators can be further miniaturized down to a few tens of micron using diamond turning \cite{Grudinin2006b}. For example, a 80 $\mu$m-diameter, 10 $\mu$m thick CaF$_{2}$ disc would possess $m_{\mathrm{eff}} = 90$ $\mathrm{ng}$ and $\Omega_{\mathrm{m}} = 63$ $\mathrm{MHz}$. If $Q_{\mathrm{m}} = 10^{7}$ can be reached, the power required to cool such a device from $1.6$ K to $T=\frac{\hbar\Omega_{\mathrm{m}}}{k_{\mathrm{B}}}$ is only 100 $\mu$W in the resolved-sideband regime \cite{Schliesser2008}. Heating due to absorbtion is likely to be totally negligible considering the optical quality of the crystal. \newline
Beyond cooling, in the regime of the parametric instability, this system may serve as a low-noise photonic RF oscillator \cite{Vahala2008}, due to its high Q-factor. The high attained sideband factor is of interest for a variety of studies, e.g. back-action evading measurements using a dual pump \cite{Braginsky1992}. In addition, the regime of strong optomechanical coupling \cite{Groeblacher2009} can be accessed with only moderate coupling rates. The crystalline resonators can also operate in the novel regime where despite high mechanical Q-factor ($>1000$) the mechanical dissipation rate $\Gamma_{\mathrm{m}}$ can exceed $\kappa$. Pumping the cavity on the upper sideband would lead to optical gain and eventually the emission of a Stokes wave at a frequency of $\omega-\Omega_{\mathrm{m}}.$ In this regime, the analogon of an optomechanical Brillouin laser can be realized, as recently analyzed \cite{Grudinin2010}. 

We thank A. Matkso for valuable comments. This work was supported by the EU program MINOS and the Max Planck Institute of Quantum Optics.

\bibliographystyle{apsrev}

\end{document}